# Towards a Formal Approach for Detection of Vulnerabilities in the Android Permissions System

Amirhosein Sayyadabdi
Faculty of Computer Engineering
University of Isfahan
Isfahan, Iran
ahsa@eng.ui.ac.ir

Behrouz Tork Ladani
Faculty of Computer Engineering
University of Isfahan
Isfahan, Iran
ladani@eng.ui.ac.ir

Bahman Zamani
MDSE Research Group
Faculty of Computer Engineering
University of Isfahan
Isfahan, Iran
zamani@eng.ui.ac.ir

*Abstract*— **Android is a widely used operating system that employs a permission-based access control model. The Android Permissions System (APS) is responsible for mediating application resource requests. APS is a critical component of the Android security mechanism; hence, a failure in the design of APS can potentially lead to vulnerabilities that grant unauthorized access to resources by malicious applications. In this paper, we present a formal approach for modeling and verifying the security properties of APS. We demonstrate the usability of the proposed approach by showcasing the detection of a well-known vulnerability found in Android's custom permissions.**

## I. INTRODUCTION

Android is the dominant end-user operating system currently available on the market [1]. Different devices, e.g., smart home appliances, TVs, and mobile phones employ Android to manage their hardware and software resources and to serve their users. Android uses the concept of permissions to manage the users' access to resources. Applications can access resources if and only if they obtain appropriate permissions either via users' consent or by declaring permissions in their source code.

The Android Permission System (APS) is a critical component of Android's security mechanism that protects private user data and sensitive system resources [2]. A flaw in the design or implementation of APS can result in a violation of the security of Android and potentially leads to critical vulnerabilities [3]. APS aims to prevent unauthorized access to sensitive resources and users' private data in Android [4]. The Android permissions system has a set of operations that applications use for gaining access to resources. The complexity of the operations and their implementations by the Android platform can lead to potential security flaws and vulnerabilities.

There are many cases so far that researchers have found permission-related issues in Android, and in response, Google has introduced security patches [3]. Due to the complexity of APS, in some cases, the patches have not been successful in terminating the vulnerabilities, and the same vulnerabilities with different attack flows remained exploitable [3], [5].

APS continuously evolves along with the Android platform [2]. Attackers target the shortcomings of APS to their advantage [2]. In practice, malicious applications can gain unauthorized access to system resources and users' private data because of the APS issues. Formally specifying the access control mechanism in Android gives a deeper understanding of the operating system, and it allows us to perform a thorough investigation of APS and analyze its security [6]. Many APS issues are the design flaws that require "system-wide reasoning," and conventional methods such as testing and static analysis are not very useful in detecting the flaws because those methods are more appropriate for detecting issues in individual components [4].

The main problem that we are trying to address is to validate the security of APS through the formal specification of the behavior of APS and verification of the security properties of APS. This paper presents a behavioral model of APS and specifies a general security property that captures the essence of a known vulnerability. We present a simple model that captures APS relevant behaviors, which helps detect a design flaw of custom permissions. We then specify a set of operations and define a security property that implies the permission vulnerability. Finally, we verify the property using the Temporal Logic Checker (TLC).

The contributions of the paper can be summarized as follows: (1) presenting a formal approach for modeling the APS, (2) modeling the behavior of APS as a case study for the detection of vulnerabilities, and (3) verifying a security property that reveals a vulnerability of custom permissions.

The rest of the paper is organized as follows. In Section II, we briefly introduce the required background information. In Section III, we discuss the related papers. In Section IV, we present our formal approach, and in Section V, we describe our implementation of a relevant case study, and we evaluate the usefulness of our approach. Finally, Section VI concludes the paper.

## II. BACKGROUND AND MOTIVATION

APS has evolved considerably compared to its initial version, e.g., runtime permissions were introduced in Android 6 [1]. Users now install applications without having to consent to any specific permissions, and APS is responsible for prompting the user and asking for permission whenever the application requests access.

Android applications cannot directly request for scarce resources (i.e., sensitive resources). Instead, an application first asks the Binder Inter-Process Communication (Binder IPC) module in the Security-Enhanced Linux (SELinux) kernel (Figure 1). The Binder IPC then carries out the task of communicating with the Android API libraries until the request for the resource is granted.

The set of permissions used to be static and predefined by Android. However, now applications are allowed to define custom permissions for their internal resources, such as databases and content providers. There was a design oversight in the custom permissions of Android 6 that resulted from the ambiguity of precedence between different custom permissions with the same name that have different protection levels, and the order of installation of the applications determines the actual permission check [4].

Android 10 has made another leap forward and introduced the concept of Non-Binary Context Dependent (NBCD) permissions. APS is now responsible for taking the current context of the device into account as well. Context is a general term for any specific situation on the device. For example, APS should consider whether an app is running on the mobile screen and check whether an app has used its permissions for a specific period.

Android 11 has introduced One-Time permissions, which are particular types of temporal NBCDs. The user is no longer forced to either accept or decline a permission request in perpetuity; instead (s)he can allow the app to access the resource only once.

The evolution of APS has raised the question of users' convenience at the cost of security violations by malicious applications. Researchers have attempted and succeeded at formally specifying and verifying the security of APS before Android 11, but Google keeps adding new features to APS in almost every major release. Therefore, it would be helpful if we could pave the way for researchers who intend to analyze the effects of tweaking the APS model through modification or the addition of features.

Formal methods have been successful in modeling the behavior of APS, in specification and verification of the security properties of APS, and the detection of flaws and vulnerabilities in APS [3], [4], [6]–[11].

The recent evolution of Android mandates system-wide reasoning of the new features of APS, which is a nontrivial problem because it requires a comprehensive introduction of temporal security properties and verification of those properties. Therefore, in this paper, we present a comprehensive formal approach for analyzing the security of APS, which is presented in Section IV.

The motivation for this work is to support the constant evolution of APS and to facilitate the process of formally specifying the security properties of APS and verifying the security properties via model checking.

The prevalence of Android as the world's most widely-deployed, end-user-focused operating system has led to the introduction of new features and capabilities in Android. The new features of APS are not trivial to specify with conventional methods and approaches that previous works have provided. Existing approaches fail to support the evolution and upgrades of the permissions system because of the complexity of the new features.

III. RELATED WORK

Formal methods have been applied successfully in the analysis and verification of the security aspects of APS [3], [4], [6], [12]. Due to the rapid evolution rate of Android as the predominant mobile operating system, the permissions system also evolved to provide the users with ease of use and enhanced security concerning sensitive resources such as the device's microphone and camera. The challenge is that sometimes the users' convenience comes at the cost of compromised system security.

Tuncay et al. [3] proposed a new modular design named Cusper for the Android permission model to address the shortcomings of the APS. Cusper separates the management of system and custom permissions. The correctness of Cusper is validated by: (1) Introducing the first formal model of Android runtime permissions, (2) Extending it to describe Cusper, and (3) Formally verifying the required properties. Cusper is implemented in Android to prove its practicality.

Talegaon et al. [13], [14] took on the path of formal methods as well, but they have not employed automatic verification (e.g., via model checking). Instead, the authors relied on testing techniques based on the specification that they had provided.

Bagheri et al. [4] pointed out that prior works on the APS security analysis had primarily focused on careful, manual scrutiny. The authors provided a behavioral model of Android in terms of architectural-level operations via system-level reasoning.

We categorized the literature based on six criteria:

1. Incorporation of the APS source code in the model,
2. Consideration of dynamic permissions,
3. Formal language used for modeling,
4. Consideration of temporal permissions,
5. Evaluation method: (1) testing, (2) simulation, (3) formal proofs, (4) model checking,
6. Android version.

Table I presents a brief taxonomy of the related work. We have selected a group of papers that are related to our work in terms of the solution techniques, the evaluation methods, and the APS features covered in the study.

Complex features such as One-Time Permissions (introduced in Android 11) and Non-Binary Context-Dependent Permissions (introduced in Android 10) [1] have not been formally studied yet. As of Android 10, the APS functionalities have moved into a separate package named "PermissionController," therefore, APS is no longer a set of rules that should always hold, and it is more beneficial to model the APS as an independent state machine.

A critical drawback of previous works is the inability to support the development of APS as a component-based system. Since APS is a complex system, it is best to decompose it into smaller subsystems. For example, the principles of component-based software design [15] can come in handy when modeling the different parts of APS; however, these principles have not been used extensively by previous works.

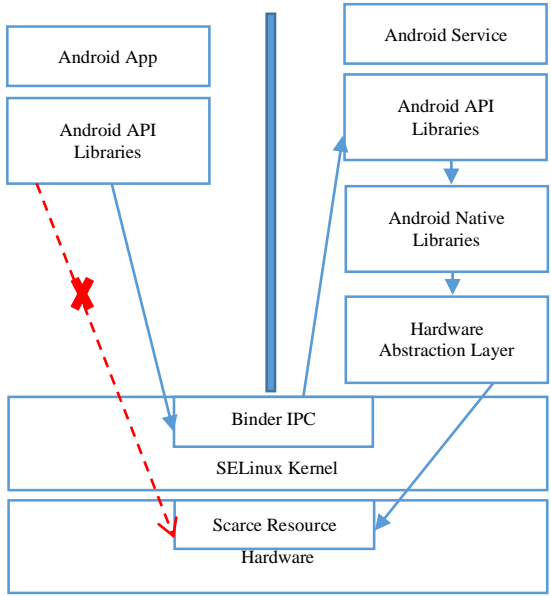

Fig. 1. The Android system model for accessing resources [16]

TABLE I. A TAXONOMY OF THE RELATED WORK.

| Work | Based on APS Source Code | Dynamic Permissions | Language | Temporal Properties | Test | Model Checking | Simulation | Proof | Android Version |
|---|---|---|---|---|---|---|---|---|---|
| Talegaon et al. [13], [14] | ■ | ■ | - | □ | ■ | □ | □ | □ | 5 |
| Almomani et al. [10] | □ | ■ | - | □ | □ | □ | □ | □ | 11 |
| Alepis et al. [17] | □ | □ | - | □ | □ | □ | □ | □ | 6 |
| Tuncay et al. [3] | ■ | ■ | Alloy | □ | ■ | ■ | □ | ■ | 6 |
| Sadeghi et al. [18] | □ | ■ | TLA+ | ■ | ■ | ■ | □ | □ | ≥ 6 |
| Betarte et al. [19] | □ | □ | Coq | ■ | ■ | □ | □ | ■ | 6 |
| Bagheri et al. [4], [20] | □ | ■ | Alloy | □ | □ | ■ | □ | □ | 6 |
| Betarte et al. [21] | □ | □ | Coq | □ | □ | □ | □ | ■ | 6 |
| Betarte et al. [22] | □ | □ | Coq | ■ | ■ | □ | □ | ■ | 6 |
| X. He [23] | □ | ■ | Petri Nets | □ | □ | □ | ■ | □ | 6 |
| Sadeghi [24] | □ | ■ | TLA+ | ■ | ■ | ■ | □ | □ | 6 |
| Betarte et al. [9] | □ | □ | Coq | ■ | □ | □ | □ | □ | 5 |
| Schmerl et al. [11] | □ | □ | Acme | □ | □ | ■ | □ | □ | 5 |
| Armando et al. [25] | □ | □ | - | □ | □ | □ | □ | ■ | < 5 |
| Fragkaki et al. [26] | □ | □ | - | □ | ■ | □ | □ | ■ | 2 |

Legend: ■ Covered  □ Not covered  - Not stated

## IV. THE PROPOSED APPROACH

An overview of our approach is depicted in Figure 2. The procedure begins by analyzing the source code of Android. The analysis phase results will form a set of interfaces that Android applications use to interact with APS. In the case of ambiguities in the behavior of APS, the source code of Android will be built and deployed to observe the actual behavior using the system logs whenever necessary.

We present a general approach that is appropriate for supporting the constant changes that are introduced into Android. Previous works fall short of providing such a general approach because the recent significant changes in Android 10 and 11 have occurred in only a few years. The system model can then be specified in TLA+, which serves as a reference for system behavior. Finally, the results of the model checker and violated properties will be used to detect flaws and vulnerabilities. As the threat model for this paper, we consider a malicious developer (i.e., attacker) who may distribute malicious applications on public stores such as Google Play. The attacker is aware that an uninformed developer has declared a privileged custom permission with a specific name.

TLA+ is a formal language for specifying digital systems and complex models. TLA+ supports the modeling of large, complex systems with hierarchical decomposition. Safety properties, liveness properties, and fairness properties are expressible in TLA+, and the TLC model checker can verify the specified properties. Every TLA+ specification consists of an initial state and a formula that specifies all the possible next states. The properties are expressible in TLA+ as well. The TLC model checker then executes all possible paths and checks whether or not the specified properties are satisfied. The flexibility and expressiveness of TLA+ makes it a convenient choice for specifying APS, which is a complex subsystem of Android.

Figure 3 presents a formal specification of the Android permissions system in TLA+. A set of applications are allowed to perform three operations: (1) get installed before the other application, (2) ask for permission, and (3) be granted the requested permissions.

Analysis of the source code plays an essential role in our approach because it contributes to the model's fidelity by ensuring that the model adheres to the actual behaviors of APS. In the meantime, the TLA+ language helps abstract the unnecessary implementation details of the source code. The security properties can be extracted from three sources: official documentation, related papers that model the permissions system, and the source code of Android. It is important to note that our approach does not rely on security properties that are based on implementation or vulnerability details.

We assume that two different applications define two custom permissions with the same name. One application is malware, which defines a custom permission with a "normal" protection level. The other application is the victim, and it also defines a custom permission with the same name but with a "dangerous" protection level. As a result of the vulnerability in APS custom permissions [4], the malware is allowed to access the resource that is protected by the victim application because it was installed before the victim.

Our approach is based on studying and observing the behavior of APS. We analyze the source code of Android to identify the interfaces along with their inputs and outputs. We then investigate the interfaces to discover the relationships between APS and other system components and applications. Finally, we present a behavioral model that captures the essence of the permissions system in a verifiable manner. Figure 4 presents our method for verifying the security properties of the Android permissions system via model checking.

The specification in Figure 3 is a representation of APS as an independent state machine that interacts with applications. The specification is available on GitHub [27]. It consists of an initial state for a group of applications, and each application can perform several operations in any possible order. The operations are specified as atomic actions in TLA⁺ that transitions the system from its current state to the next state. This method for defining actions is most useful when combined with the hierarchical decomposition technique, which can lead to a component-based model that is also extensible.

Our work takes advantage of the techniques of Component-Based Software design [15]. We present a new approach for modeling and verifying the Android permissions system by investigating the source code of Android along with the official documentation and executing the source code of Android, which can help resolve any possible ambiguities in the documentation. We also take advantage of specifying security properties at a high level in a general manner, which can help detect unknown vulnerabilities.

We take the Android's source code, the official documentation, and the list of known vulnerabilities to detect the essential features of APS. We specify the security properties as TLA⁺ formulas that the TLC model checker would check. The results of the model checking process are then investigated for the detection of violated properties. The sequence of states that lead to a violation can then be used to identify the events that caused the violation, which can help detect flaws.

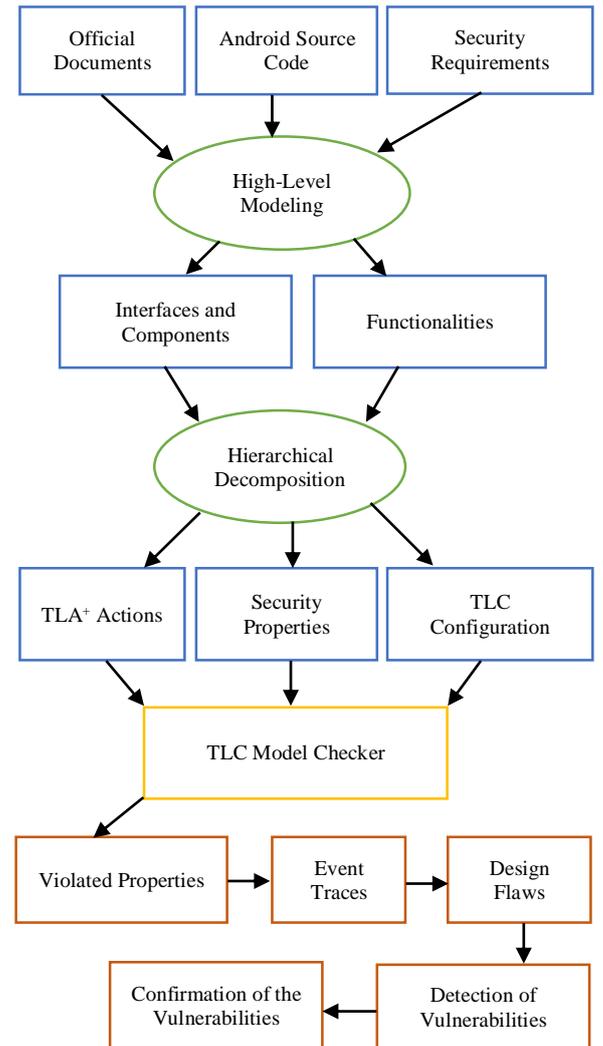

Fig. 3. The TLA⁺ specification of the behavior of APS

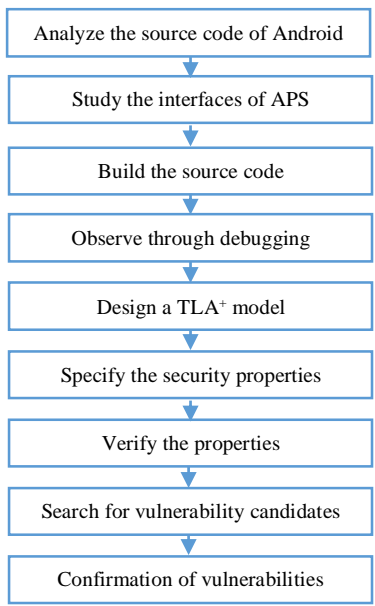

Fig. 2. The steps of our formal approach

Fig. 4. Our approach for verification of security properties

## V. IMPLEMENTATION AND EVALUATION

To showcase our approach's effectiveness and soundness, we have implemented a proof-of-concept by verifying a case study of a known vulnerability in custom permissions in Android 6. We modeled the APS in TLA$^+$, and we specified a general security property that should be satisfied by the APS model. We specify the security as the following statement (Figure 5):

*ApsConsistent = "No request for a normal permission may lead to obtaining a dangerous permission."*

$$ApsConsistent \triangleq \neg \exists\, r \in APP : \wedge\ askedPerms[r] = \text{``NOR''}$$
$$\wedge\ grantedPerms[r] = \text{``DAN''}$$

Fig. 5. A security property of the permissions system

If the *ApsConsistent* property gets violated due to an APS operation or decision, then the system's security is also considered violated. As a result, the traces of the TLC model checker will show the exact order of operations that lead to the violation, which can help identify the original vulnerability.

We show that our approach for formal specification and verification of the security of APS is useful and sound for detecting flaws in the design of a basic permissions system. Our implementation of the basic APS model yields an easy-to-read specification that helps detect a known vulnerability in the Android's custom permissions.

It is important to note that the security property *ApsConsistent* is a general property that does not imply any information about any vulnerabilities or design flaws. It is a prominent security property that one would expect of APS. The TLC model checker instead will come in handy, and it shows us how the property got violated. The TLC's explanation helps detect any possible source of error in the permissions system, which is the benefit of modeling the APS as an independent entity that facilitates the verification of simple security properties against the specification.

We have specified a general property (*ApsConsistent*) that does not convey information about the vulnerability. However, with the help of the TLC model checker and analyzing the sequence of events, we can detect the vulnerability in APS, which is an advantage of specifying the system's behavior as an independent state machine in a formal language. The reason is that the model checker can traverse through all possible behaviors of the specified system and check whether or not a set of specified properties (security properties) are satisfied. The model checking results then may help detect previously unknown flaws and vulnerabilities because the properties are agnostic about the details of the vulnerabilities.

An essential advantage of our approach is the technique for resolving the ambiguities in either the Android's source code or the official documentation. We employ the technique of observation-based modeling by building the source code of Android, which allows the debugging and the tweaking of the source code and logging of helpful information that may help resolve the ambiguities.

Another approach would be to neglect the APS ambiguities, but our approach is designed to produce a model that adheres to the underlying system. Observing the phenomena being modeled is a useful approach for gaining a more profound understanding of the ongoing events in the environment, resulting in a useful model.

We demonstrate the applicability of our approach to detecting vulnerabilities without having to specify the vulnerabilities beforehand, which is particularly useful for dealing with unknown flaws because the TLC model checker helps by checking all possible scenarios, which could lead to the detection of unknown vulnerabilities. The generality of properties is a key feature of our approach.

The TLC model checker traverses all possible sequences of states that result from applying the APS operations. TLC then checks whether the specified security property holds over all possible paths. If a property gets violated, then TLC will report an error (Figure 6 and Figure 7). TLC also provides a statistical result based on the total number of reached states and the diameter of the traversal graph.

The execution trace reported by TLC shows the order of operations that have taken place until the security property *ApsConsistent* is violated. The security property is specified as a safety property, which should hold in all possible model executions. Specifying security properties as safety properties is a useful way of verifying the security of APS.

The advantage of formally specifying the permissions system in a formal language is that it provides a deeper understanding of the existing system and helps with performing a security analysis of the permissions subsystem of Android responsible for managing applications' access to resources [6].

Complex systems such as APS require system-wide reasoning to ensure their correctness. While modeling the permissions system as a state machine can be costly and complex, it can still be useful for analyzing the desired security properties of APS and detecting the flaws that otherwise could not be detected. The comprehensiveness results from exhaustive analysis through verification via model checking a security property that is expected to be true in all possible system behaviors.

While software testing is a common practice in the industry, it is not entirely adequate when verifying the security of critical system components such as APS [3]. We chose the TLA$^+$ language to model the APS' behavior because it facilitates a comprehensive formal analysis of the security of such a complex system.

Complex systems require system-wide reasoning concerning their correctness; therefore, we modeled the permissions system as a high-level abstract entity that performs several operations in the operating system and interacts with applications. The results led to the detection of a known custom permission vulnerability in Android 6, which shows the effectiveness of our approach.

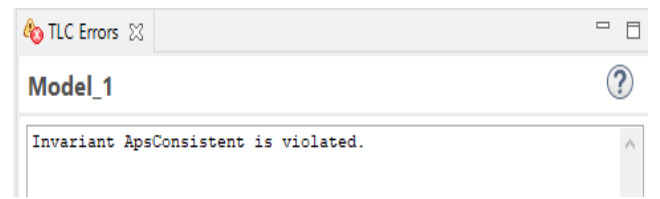

Fig. 6. TLC report on the violation of the security property

The TLA⁺ language supports the concept of a theorem (Figure 8), which is a statement that the TLC model checker can check effectively. After the successful termination of TLC with no reported errors, we can assume that the theorem is correct and there are no violations of the specified properties.

There is also the option of specifying the system using the PlusCal algorithm language. PlusCal is easier to read and resembles the syntax of programming languages. Specifications written in PlusCal will be automatically translated into TLA⁺, which can also be used to feed the TLC model checker because the security properties are still written in TLA⁺.

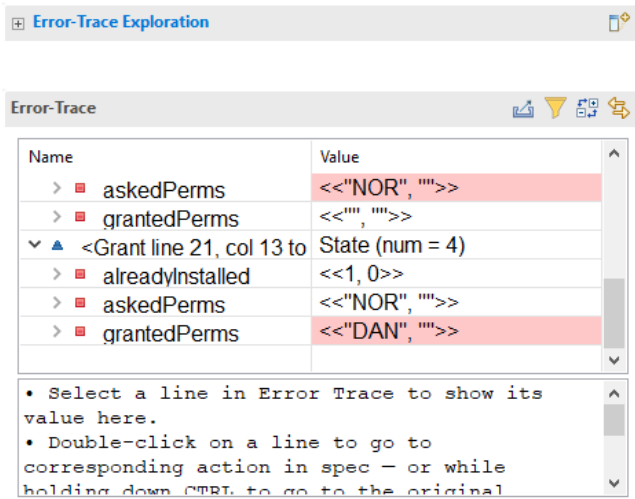

Fig. 7. An execution trace reported by TLC

$$ApsSpec \triangleq ApsInit \wedge \Box[ApsNext]_{grantedPerms}$$

$$\text{THEOREM } ApsSpec \Rightarrow \Box(ApsTypeOK \wedge ApsConsistent)$$

Fig. 8. A theorem to be verified using the TLC model checker

The process of extending the APS specification includes the identification of other operations that are implemented by the Android platform. Since TLA⁺ supports hierarchical decomposition, we can model separate parts of APS as independent components.

By designing the basic model of APS as a component-based artifact, we achieve the goal of extensibility that will be useful for upgrading the model to adhere to the next features of Android.

The principles of component-based software design [15] are applicable as well because we can implement the same ideas in our formal model. The ultimate advantage of designing a component-based model of APS is that we can also build a framework that facilitates the modeling of future versions of Android. Our approach is designed to support the future evolution and upgrades of the Android operating system, which is a distinguishing feature compared with other existing approaches.

## VI. CONCLUSION AND FUTURE WORK

The Android permissions system (APS) is a critical component of the Android operating system. The complexity of APS mandates system-wide reasoning and investigation of the underlying platform. Presenting a formal approach for modeling and verifying APS is useful for supporting the future versions of Android and verifying its security properties.

The complexity of the permissions system now requires support for extending the model in the future. One way of supporting the extensibility of the model is to follow well-known software engineering techniques, such as component-based design principles, which we employed in our approach via the support of the TLA⁺ formal language for hierarchical decomposition.

We presented a new approach for formal specification and verification of the security properties of APS. We also presented a basic model of custom permissions in APS. We then used the model to verify a general security property, which led to the detection of an existing vulnerability.

We showed that our approach yields a basic model that can help detect a known vulnerability by verifying a general security property that did not convey any information about the vulnerability itself. We showed the effectiveness of our approach in finding flaws and detecting design vulnerabilities in the permissions system, which is useful because it can also detect unknown vulnerabilities since the model checker traverses all possible behaviors of the specification.

Our formal approach is designed to result in a valid, verifiable, and extensible model of APS, which is the key distinguishing feature of our new approach compared with other existing formal approaches.

Extending the basic model of APS to support the newest features of Android would be a fit candidate for the next step in fully implementing our formal approach to the verification of the security properties of APS.

The application of component-based design principles will also be an important aspect of future attempts to model the complete behavior of APS. Designing the basic model as an extensible model would be useful for supporting the future versions of Android because components can then be modeled by independent teams and incorporated into the basic model, which also helps with error detection because of the segregation of components' responsibilities.

Scrutiny of the APS source code will also yield a better understanding of the actual behavior of APS, which can help resolve any possible ambiguities in the official documentation of the Android operating system.

One possible future direction is to design a static analysis tool that automatically investigates the source code of Android and detects the interfaces of APS and their inputs and outputs. This tool will produce results that will be useful in the modeling process. The procedure of detecting APS interfaces is currently a manual task in our approach, which could benefit from being automated using a static analysis tool.

ACKNOWLEDGMENT

This work is partially based upon research funded by Iran National Science Foundation (INSF) under project No. 4003042.